\documentclass[11pt]{article}
\usepackage{amssymb}

\newtheorem{lemma}{Lemma}
\newenvironment{proof}[1][Proof]{\textbf{#1.} }{\ \rule{0.5em}{0.5em}}
\evensidemargin\oddsidemargin
\begin{document}

\bigskip

\begin{center}
{\LARGE The role of infrared divergence for decoherence}

\bigskip

{\large Joachim Kupsch\footnote{{e-mail: kupsch@physik.uni-kl.de}}}

{\large Fachbereich Physik, Universit\"at Kaiserslautern\\
D-67653 Kaiserslautern, Germany}

\medskip
\end{center}

\begin{abstract}
Continuous and discrete superselection rules induced by the interaction with
the environment are investigated for a class of exactly soluble Hamiltonian
models. The environment is given by a Boson field. Stable superselection sectors
emerge if and only if the low frequences dominate and the ground state of the Boson
field disappears due to infrared divergence. The models allow uniform
estimates of all transition matrix elements between different superselection
sectors.
\end{abstract}
                    
\section{Introduction}

Superselection rules are the basis for the emergence of classical physics
within quantum theory. But despite of the great progress in understanding
superselection rules, see e.g. \cite{Wightman:1995}, quantum mechanics and
quantum field theory do not provide enough superselection rules to infer the
classical probability of ``facts'' from quantum probability. This problem is
most often discussed in the context of measurement of quantum mechanical
objects. In an important paper about the process of measurement Hepp \cite
{Hepp:1972} has presented a class of models for which the dynamics induces
superselection sectors. Hepp starts with a very large algebra of observables
-- essentially all observables with the exception of the ``observables at
infinity'' which constitute an a priory set of superselection rules -- and
the superselection sectors emerge in the weak operator convergence. But it
has soon been realized that the algebra of observables, which is relevant
for the understanding of the process of measurement \cite{Emch:1972a} \cite
{Araki:1980} and, more generally for the understanding of the classical
appearance of the world \cite{Zurek:1982} \cite{Joos/Zeh:1985} \cite
{GJKKSZ:1996} can be severely restricted. Then strong or even uniform
operator convergence is possible.

In this paper results of Chap.7 of the book \cite{GJKKSZ:1996} and of the
article \cite{Kupsch:1998} are extended. After a short introduction to
superselection rules and the dynamics of subsystems we prove in Sect.\ref
{models} that uniform operator estimates are possible also for continuous
superselection rules induced by the environment. In Sect.\ref{field} we
investigate a class of Hamiltonian models with an environment given by a
Boson field. The restriction to the Boson sector corresponds to a van Hove
model \cite{Hove:1952}. As the main result of the paper we prove for this
class of models:\newline
-- The superselection sectors are induced by the infrared contributions of
the Boson field.\newline
-- The superselection sectors are stable for $t\rightarrow \infty $ if and
only if the Boson field is infrared divergent.

This type of infrared divergence has been studied by Schroer \cite
{Schroer:1963} more than thirty years ago. The Boson field is still defined
on the Fock space but the ground state of the Boson field disappears in the
continuum.

\section{Induced superselection rules\label{ssr}}

We start with a few mathematical notations. Let $\mathcal{H}$ be a separable
Hilbert space, then the following spaces of linear operators are used. 
\newline
$\mathcal{B}(\mathcal{H})$: The linear space of all bounded operators $A$
with the operator norm $\Vert A\Vert $.\newline
$\mathcal{T}(\mathcal{H})$: The linear space of all nuclear operators $A$
with the trace norm $\Vert A\Vert _{1}=\mathrm{tr}\sqrt{A^{+}A}$.\newline
$\mathcal{D}(\mathcal{H})$: The set of all positive nuclear operators $W$
with a normalized trace, $\mathrm{tr}\,W=1$.

We consider standard quantum mechanics and quantum field theory where any
state of a quantum system is represented by a statistical operator $W\in 
\mathcal{D}(\mathcal{H})$ - the rank one projection operators thereby
correspond to the pure states - and any bounded observable is represented by
an operator $A\in \mathcal{B}(\mathcal{H})$. Without additional knowledge
about the structure of the system we have to assume that the set of all
states corresponds to $\mathcal{D}(\mathcal{H})$, and the operator algebra
of all (bounded) observables coincides with $\mathcal{B}(\mathcal{H})$. In
quantum field theory the superposition principle is partially restricted to
superselection sectors, see e.g. \cite{Wightman:1995}. The projection
operators onto the superselection sectors commute with all observables of
the theory: they are classical observables. But there remains an essential
problem for the understanding of the classical appearance of the world: Only
very few superselection rules can be found in quantum mechanics and quantum
field theory. A possible solution is the emergence of superselection rules
due to decoherence caused by the dynamics.

Let $A=A(0)\rightarrow A(t)=\mathcal{T}_{t}(A)\in \mathcal{B}(\mathcal{H})$
denote the dynamics in the Heisenberg picture. If there exists a family of
projection operators $\left\{ P_{m},m\in \mathbf{M}\right\} $ with the
properties $P_{m}P_{n}=0$ for $m\neq n$ and $\sum_{n}P_{n}=I$, such that
transition matrix elements $\left( f\mid A(t)g\right) $ between different
sectors $f\in \mathcal{H}_{m}=P_{m}\mathcal{H},\,\,g\in \mathcal{H}_{n}=P_{n}%
\mathcal{H},\,\,m\neq n$, vanish for all observables $A\in \mathcal{B}(%
\mathcal{H})$ for $t\rightarrow \infty $, the subspaces $\mathcal{H}%
_{m}=P_{m}\mathcal{H},\,m\in \mathbf{M,}$ are denoted as\textit{\ }%
superselection sectors induced by the dynamics $\mathcal{T}_{t}$.

This definition can be applied to the Hamiltonian dynamics $%
A=A(0)\rightarrow A(t)=\mathcal{T}_{t}(A):=U^{+}(t)AU(t)$ where $U(t)=\exp
(-iHt)$ is the unitary group generated by the Hamiltonian $H$. As a simple
example we consider a Hamiltonian $H$ on the Hilbert space $\mathcal{H}=%
\mathcal{H}_{0}\oplus \mathcal{H}_{1}$ with one bound state at energy $E_{0}$
in the 1-dim. subspace $\mathcal{H}_{0}$ and with an absolutely continuous
spectrum in the subspace $\mathcal{H}_{1}$. Then for $f_{0}\in \mathcal{H}%
_{0}$ and $f_{1}\in \mathcal{H}_{1}$ with $\left\| f_{0,1}\right\| =1$ we
calculate $\left( f_{0}\mid A(t)f_{1}\right) =e^{iE_{0}t}\left(
A^{+}(0)f_{0}\mid U(t)f_{1}\right) \rightarrow 0$, since $U(t)f_{1}$
converges weakly to zero. The subspaces $\mathcal{H}_{0}$ and $\mathcal{H}%
_{1}$ are therefore induced superselection sectors of the Hamiltonian
dynamics. If $P_{0,1}$ denote the projection operators onto the subspaces $%
\mathcal{H}_{0,1}$ then the off-diagonal part $P_{0}A(t)P_{1}$ converges in
the weak operator norm to zero. But neither strong nor, a fortiori, uniform
convergence holds for $P_{0}A(t)P_{1}$ unless $P_{0}A(t)P_{1}\equiv 0$. More
refined examples have been given by Hepp \cite{Hepp:1972}. Thereby an
essential consequence of the Hamiltonian time evolution or any other
automorphic time evolution is the restriction to a weak operator
convergence. Moreover, as has been emphasized by Bell \cite{Bell:1975}, the
time scale can be arbitrarily long, such that the practical use of such
models is questionable.

A strong or even uniform suppression of the off-diagonal matrix elements of
all observables can be obtained by the restriction to a subsystem \cite
{Emch:1972a} \cite{Araki:1980} \cite{Zurek:1982}. In the following we
consider an open system, i.e. a system $\mathcal{S}$ which interacts with an
environment $\mathcal{E}$, such that the total system $\mathcal{S}\times 
\mathcal{E}$ satisfies the usual Hamiltonian dynamics. The Hilbert space $%
\mathcal{H}_{S\times E}$ of the total system $\mathcal{S}\times \mathcal{E}$
is the tensor space $\mathcal{H}_{S}\otimes \mathcal{H}_{E}$ of the Hilbert
spaces for $\mathcal{S}$ and for $\mathcal{E}$. If the state of the total
system is $W\in \mathcal{D}(\mathcal{H}_{S+E})$, then the state of the
subsystem is given by the reduced statistical operator $\rho =\mathrm{tr}%
_{E}W\in \mathcal{D}(\mathcal{H}_{S})$. The dynamics of the states of the
total system $W\in \mathcal{D}(\mathcal{H}_{S\times E})\rightarrow
W(t)=U(t)W(0)U^{+}(t)\in \mathcal{D}(\mathcal{H}_{S\times E})$ with the
unitary group $U(t)=\exp (-iHt)$, generated by the total Hamiltonian $H$,
yields the dynamics of the statistical operator $\rho (t)=\mathrm{tr}%
_{E}\,U(t)W(0)U^{+}(t)\in \mathcal{D}(\mathcal{H}_{S})$ of the subsystem $%
\mathcal{S}$. In the following we assume that the initial state factorizes $%
W=\rho \otimes \omega $ with $\rho \in \mathcal{D}(\mathcal{H}_{S})$ and a
fixed reference state $\omega \in \mathcal{D}(\mathcal{H}_{E})$ of the
environment. Then the dynamics in the Heisenberg picture of the system $%
\mathcal{S}$ is easily calculated as 
\begin{equation}
A\in \mathcal{B}(\mathcal{H}_{S})\rightarrow A(t)=\mathcal{T}_{t}(A):=%
\mathrm{tr}_{E}\,U^{+}(t)(A\otimes I_{E})U(t)\omega \in \mathcal{B}(\mathcal{%
H}_{S}).  \label{d.7}
\end{equation}
Before we investigate induced superselection sectors we generalize the
definition given above to the case of continuous superselection sectors. The
finite or countable set of projection operators $\left\{ P_{m},m\in \mathbf{M%
}\right\} $ is substituted by a strongly continuous family of projection
operators $P(\Delta )$ indexed by measurable subsets $\Delta \subset \mathbb{R}$%
, see e.g. \cite{Piron:1969} or \cite{Araki:1980}. These projection
operators have to satisfy 
\begin{equation}
\left\{ 
\begin{array}{l}
P(\Delta _{1}\cup \Delta _{2})=P(\Delta _{1})+P(\Delta _{2})\;\mathrm{and}%
\;P(\Delta _{1})P(\Delta _{2})=0\;\mathrm{if}\;\,\Delta _{1}\cap \Delta
_{2}=\emptyset \\ 
P(\emptyset )=0,\;P(\mathbb{R})=1.
\end{array}
\right.  \label{obs.8}
\end{equation}
If we chose for $\left\{ P(\Delta ),\,\Delta \subset \mathbb{R}\right\} $ a
general (right continuous) spectral family, the case of discrete
superselection rules is included in (\ref{obs.8}).

The dynamics of the total system induces superselection rules in the system $%
\mathcal{S}$ if there exists a right continuous family of projection
operators (\ref{obs.8}) $\left\{ P_{S}(\Delta )\mid \Delta \subset \mathbb{R}%
\right\} $ defined on the Hilbert space $\mathcal{H}_{S}$, such that the
off-diagonal contributions of all statistical operators of the system $%
\mathcal{S}$ vanish for $t\rightarrow \infty $, i.e. $P(\Delta _{1})\rho
(t)P(\Delta _{2})\rightarrow 0$\ if\ $\,t\rightarrow \infty $\ and\ $\Delta
_{1}\cap \Delta _{2}=\emptyset $, or in the Heisenberg picture, $%
P_{S}(\Delta _{1})A(t)P_{S}(\Delta _{2})\rightarrow 0\;\mathrm{if}%
\;\,t\rightarrow \infty \;\mathrm{and}\;\Delta _{1}\cap \Delta
_{2}=\emptyset $ for all observables $A\in \mathcal{B}(\mathcal{H}_{S})$.

\section{Soluble models\label{models}}

In the following we present models for which the Hamiltonian of the total
system provides a family of projection operators $\left\{ P_{S}(\Delta
),\,\Delta \subset \mathbb{R}\right\} $ on $\mathcal{H}_{S}$ such that the
off-diagonal elements of any bounded observable of the system $\mathcal{S}$
can be estimated with the operator norm. We derive a uniform decrease 
\begin{equation}
\left\| P_{S}(\Delta _{1})A(t)P_{S}(\Delta _{2})\right\| \rightarrow 0\;%
\mathrm{if}\;t\rightarrow \infty  \label{mod.1}
\end{equation}
for arbitrary bounded observables $A\in \mathcal{B}(\mathcal{H}_{S})$ and
arbitrary disjoint closed intervals $\Delta _{1}\cap \Delta _{2}=\emptyset $.

The models have the following structure. The total Hamiltonian is defined on
the tensor space $\mathcal{H}_{S\times E}=\mathcal{H}_{S}\otimes \mathcal{H}%
_{E}$ as 
\begin{eqnarray}
H_{S\times E} &=&H_{S}\otimes I_{E}+I_{S}\otimes H_{E}+F\otimes G  \nonumber
\\
&=&\left( H_{S}-\frac{1}{2}F^{2}\right) \otimes I_{E}+\frac{1}{2}\left(
F\otimes I_{E}+I_{S}\otimes G\right) ^{2}+I_{S}\otimes \left( H_{E}-\frac{1}{%
2}G^{2}\right)  \label{mod.3}
\end{eqnarray}
where $H_{S}$ is the positive Hamiltonian of $\mathcal{S}$, $H_{E}$ is the
positive Hamiltonian of $\mathcal{E}$, and $F\otimes G$ is the interaction
potential between $\mathcal{S}$ and $\mathcal{E}$ with operators $F$ on $%
\mathcal{H}_{S}$ and $G$ on $\mathcal{H}_{E}$. To guarantee that $H_{S\times
E}$ is self-adjoint and semibounded we assume

\begin{enumerate}
\item[1)]  The operators $F$ and $F^{2}$ ($G$ and $G^{2})$ are essentially
self-adjoint on the domain of $H_{S}$ ($H_{E}$). The operators $H_{S}-\frac{1%
}{2}F^{2}$ and $H_{E}-\frac{1}{2}G^{2}$ are semibounded.
\end{enumerate}

Since $F^{2}\otimes I_{E}\pm 2F\otimes G+I_{S}\otimes G^{2}$ are positive
operators, the operator $F\otimes G$ is \newline
$\left( H_{S}\otimes I_{E}+I_{S}\otimes H_{E}\right) $-bounded with relative
bound one, and W\"{u}st's theorem, see e.g. Theorem X.14 in \cite
{Reed/Simon:1975}, implies that $H_{S\times E}$ is essentially self-adjoint
on the domain of $H_{S}\otimes I_{E}+I_{S}\otimes H_{E}$. Moreover $%
H_{S\times E}$ is obviously semibounded.

To derive induced superselection rules we need the rather severe restriction

\begin{enumerate}
\item[2)]  The operators $H_{S}$ and $F$ commute strongly, i.e. their
spectral projections commute.
\end{enumerate}

So far no model with Hamiltonian dynamics has been presented which violates
this assumption and allows the uniform estimate (\ref{mod.1}) of induced
superselection sectors. If the Hamiltonian includes a scattering potential
it is possible to abandon this assumption. But then the off-diagonal terms $%
P(\Delta _{1})A(t)P(\Delta _{2})$ decrease only in the strong operator
topology, see \cite{Kupsch:2000}.

The operator $F$ has a spectral decomposition $F=\int_{\mathbb{R}}\lambda
P_{S}(d\lambda )$ with a right continuous family of projection operators $%
P_{S}(\Delta )$ indexed by measurable subsets $\Delta \subset \mathbb{R}$. We
shall see below that exactly the projection operators of this spectral
decomposition determine the induced superselection sectors.

As a consequence of assumption 2) we have $\left[ H_{S},P_{S}(\Delta )\right]
=0$ for all intervals $\Delta \subset \mathbb{R}$. The Hamiltonian (\ref{mod.3}%
) has therefore the form $H_{S\times E}=H_{S}\otimes I_{E}+\int_{\mathbb{R}%
}P_{S}(d\lambda )\otimes \left( H_{E}+\lambda G\right) $. The operator $%
\left| G\right| =\sqrt{G^{2}}$ has the upper bound $\left| G\right| \leq
aG^{2}+(4a)^{-1}I$ with an arbitrarily small constant $a>0$. Since $G^{2}$
is $H_{E}$-bounded with relative bound $2$, the operator $G$ is $H_{E}$%
-bounded with an arbitrarily small bound. The Kato-Rellich theorem, see e.g. 
\cite{Reed/Simon:1975}, implies that the operators $H_{E}+\lambda G$ are
self-adjoint on the domain of $H_{E}$ for all $\lambda \in \mathbb{R}$. The
unitary evolution $U(t):=\exp (-iH_{S\times E}t)$ of the total system can
therefore be written as $U(t)=\left( \mathrm{e}^{-iH_{S}t}\otimes
I_{E}\right) \int dP_{S}(\lambda )\otimes \mathrm{e}^{-i\left( H_{E}+\lambda
G\right) t}$. The dynamics of the observables (\ref{d.7}) follows as 
\begin{equation}
A(t)=\mathrm{e}^{iH_{S}t}\left( \int \int \chi \left( \alpha ,\beta
;t\right) P_{S}(d\alpha )\,A\,P_{S}(d\beta )\right) \mathrm{e}^{-iH_{S}t}
\label{mod.10}
\end{equation}
with the trace 
\begin{equation}
\chi (\alpha ,\beta ;t)=\mathrm{tr}_{E}\left( \mathrm{e}^{i\left(
H_{E}+\alpha G\right) t}\mathrm{e}^{-i\left( H_{E}+\beta G\right) t}\omega
\right) .  \label{mod.11}
\end{equation}
The emergence of dynamically induced superselection rules depends on an
estimate of this trace. For the models investigated below, we obtain for a
large class of reference states $\omega $ (actually a dense set within $%
\mathcal{D}(\mathcal{H}_{E})$) the bounds 
\begin{equation}
\left| \frac{\partial ^{n}}{\partial \alpha ^{n}}\chi (\alpha ,\beta
;t)\right| \leq c\left( 1+(\alpha -\beta )^{2}\psi (t)\right) ^{-\gamma
},\;n=0,1,  \label{mod.12}
\end{equation}
with a function $\psi (t)\geq 0$ which diverges for $t\rightarrow \infty $
like a power $t^{\delta },\,0<\delta <1$, and an exponent $\gamma >0$ which
can be a large number. If $\Delta _{1}$ and $\Delta _{2}$ are intervals with
a distance $\delta >0$ then the operator norm of $P_{S}(\Delta
_{1})A(t)P_{S}(\Delta _{2})$ is estimated in the Appendix \ref{estimates} as 
\begin{equation}
\left\| P_{S}(\Delta _{1})A(t)P_{S}(\Delta _{2})\right\| \leq const\left\|
A\right\| \left( 1+\delta ^{2}\psi (t)\right) ^{-\gamma }.  \label{mod.14}
\end{equation}
For operators $F$ with a discrete spectrum $F=\sum \lambda _{n}P_{n}^{S}$
uniform norm estimates have already been derived in Sect. 7.6 of \cite
{GJKKSZ:1996}. In this case the bound with $n=1$ in (\ref{mod.12}) is
obsolete.

A simple class of explicitly soluble models which yield the estimates (\ref
{mod.12}) can be obtained under the additional assumption

\begin{enumerate}
\item[3)]  The Hamiltonian $H_{E}$ and the potential $G$ commute strongly.
The operator $G$ has an absolutely continuous spectrum.
\end{enumerate}

Such models have been investigated (for operators $F$ with a discrete
spectrum) by Araki \cite{Araki:1980} and by Zurek \cite{Zurek:1982}, see
also Sect. 7.6 of \cite{GJKKSZ:1996} and \cite{Kupsch:2000}. Under the
assumption 3) the trace (\ref{mod.11}) simplifies to $\chi (\alpha ,\beta
;t)=\mathrm{tr}_{E}\left( \mathrm{e}^{i(\alpha -\beta )Gt}\omega \right) $.
Let $G=\int_{\mathbb{R}}\lambda P_{E}(d\lambda )$ be the spectral
representation of the operator $G$. Then the measure $d\mu (\lambda ):=%
\mathrm{tr}_{E}\left( P_{E}(d\lambda )\,\omega \right) $ is absolutely
continuous with respect to the Lebesgue measure for any $\omega \in \mathcal{%
D}(\mathcal{H}_{E})$, and the function $\chi (t):=\mathrm{tr}_{E}\left( 
\mathrm{e}^{iGt}\omega \right) =\newline
\int_{\mathbb{R}}\mathrm{e}^{i\lambda t}$ $d\mu (\lambda )$ vanishes for $%
t\rightarrow \infty $. But to obtain a decrease which is effective in
sufficiently short time, we need an additional smoothness condition on $%
\omega $. This condition does not impose restrictions on the statistical
operator $\rho \in \mathcal{D}(\mathcal{H}_{S})$ of the system $\mathcal{S}$%
. We assume that $G\omega \in \mathcal{T}(\mathcal{H}_{E})$ and, moreover,
that the integral operator, which represents $\omega $ in the spectral
representation of $G$, is a sufficiently differentiable function vanishing
at the boundary points of the spectrum. Then the measure $d\mu (\lambda )=%
\mathrm{tr}_{E}\left( P_{E}(d\lambda )\,\omega \right) $ has a smooth
density, and we can derive a strong decrease of its Fourier transform $\chi
(t)$ and its derivative, $\left| \frac{d^{n}}{dt^{n}}\chi (t)\right| \leq
C_{\gamma }(1+t^{2})^{-\gamma },$\thinspace $n=0,1$, with arbitrarily large
values of $\gamma $. That implies bounds (\ref{mod.12}) with $\psi (t)=t^{2}$%
.

\section{The interaction with a Boson field\label{field}}

In this section we present a model without the restriction 3). Preliminary
results have already been reported in \cite{Kupsch:2000}. We choose a system 
$\mathcal{S}$ which satisfies the constraints 1) and 2). The environment
given by a Boson field is investigated in details below. As essential result
we derive the uniform estimates (\ref{mod.12}). Consequently the
off-diagonal elements of the operator $F$ are suppressed as given in (\ref
{mod.14}). As specific example we may consider a particle on the real line
with velocity coupling. The Hilbert space of the particle is $\mathcal{H}%
_{S}=\mathcal{L}^{2}(\mathbb{R})$. The Hamiltonian and the interaction
potential of the particle are 
\begin{equation}
H_{S}=\frac{1}{2}P^{2}\;\mathrm{and}\;F=P  \label{field.1}
\end{equation}
where $P=-i\frac{d}{dx}$ is the momentum operator of the particle. The
identity $H_{S}-\frac{1}{2}F^{2}=0$ guarantees the positivity of the first
term in (\ref{mod.3}). Decoherence then yields superselection rules for the
momentum of the particle.

As Hilbert space $\mathcal{H}_{E}$ we choose the Fock space of symmetric
tensors $\mathcal{F}(\mathcal{H}_{1})$ based on the one particle Hilbert
space $\mathcal{H}_{1}$. The inner product of $\mathcal{F}(\mathcal{H}_{1})$
is denoted by $\left( .\mid .\right) $. The Hamiltonian is generated by a
one-particle Hamilton operator $M$ on$\,\mathcal{H}_{1}$ with the following
properties

(i) $M\;$is\thinspace a\thinspace positive operator with an absolutely
continuous spectrum,

(ii) $M\;$has\thinspace an\thinspace unbounded\thinspace inverse$\,M^{-1}$.

The spectrum of $M$ is (a subset of) $\mathbb{R}_{+}$, which -- as a
consequence of the second assumption -- includes zero. The Hamiltonian of
the free field is then the derivation $H_{E}=d\Gamma (M)$ generated by $M$,
see Appendix \ref{vanHove}. Let $a^{+}(f)$ denote the creation operator of
the one-particle state $f\in \mathcal{H}_{1}$ and $a(f)=\left(
a^{+}(f)\right) ^{+}$ the corresponding annihilation operator, normalized to 
$\left[ a(f),a^{+}(g)\right] =\left( f\mid g\right) $. The interaction
potential $G$ is then chosen as the self-adjoint field operator $G=\Phi
(h):=a^{+}(h)+a(h)$, where $h\in \mathcal{H}_{1}$ satisfies the additional
constraint 
\begin{equation}
2\left\| M^{-\frac{1}{2}}h\right\| \leq 1.  \label{field.4}
\end{equation}
This constraint guarantees that $H_{E}-\frac{1}{2}\Phi ^{2}(h)$ is bounded
from below, and the Hamiltonian (\ref{mod.3}) is a well defined semibounded
operator on $\mathcal{F}(\mathcal{H}_{S\times E})$, see Appendix \ref
{vanHove}.

To derive induced superselection sectors for the observable $P$ we have to
estimate the time dependence of the traces (\ref{mod.11}) $\chi _{\alpha
\beta }(t):=\mathrm{tr}_{E}U_{\alpha \beta }(t)\omega ,\,\alpha \neq \beta ,$
where $\omega $ is the reference state of the Boson field, and the unitary
operators $U_{\alpha \beta }(t)$ are given by 
\begin{equation}
U_{\alpha \beta }(t):=\exp (iH_{\alpha }t)\exp (-iH_{\beta }t),\;\mathrm{with%
}\;H_{\alpha }=H_{E}+\alpha \Phi (h),\;\alpha ,\beta \in \mathbb{R}\mathbf{.}
\label{field.5}
\end{equation}
The Hamiltonians $H_{\alpha }$ are Hamiltonians of the van Hove model \cite
{Hove:1952}. In the Appendix \ref{vanHove} we prove the following results
for reference states $\omega $ which are finite superpositions or mixtures
of coherent states.

\begin{enumerate}
\item  If the vector $h$ also satisfies $M^{-1}h\in \mathcal{H}_{1}$ one can
use the standard methods of the van Hove model to evaluate the traces $\chi
_{\alpha \beta }(t)=\mathrm{tr}_{E}U_{\alpha \beta }(t)\omega $. These
traces do not vanish for $t\rightarrow \infty $. But one can achieve a
strong decrease which persists for some finite time interval. This period
can be arbitrarily large; but inevitably, recurrences exist.

\item  If $M^{-1}h\notin \mathcal{H}_{1}$ the low energy contribution of the
interaction potential dominates, and $\chi _{\alpha \beta }(t)$ vanishes for 
$t\rightarrow \infty $ if $\alpha \neq \beta $. If the vector $h$ satisfies
some additional regularity condition at small energies, there exists a
uniform limit $\lim_{t\rightarrow \infty }\chi _{\alpha \beta }(t)=0$ for
all $\alpha ,\beta $ with $\left| \alpha -\beta \right| \geq \delta >0$, and
zero can be approached within a short time.
\end{enumerate}

\noindent The assumption $M^{-1}h\notin \mathcal{H}_{1}$ is therefore
necessary and sufficient for the emergence of superselection rules, which
persist for $t\rightarrow \infty $. In this case the Boson field is infrared
divergent. It is still defined on the Fock space, but its ground state
disappears in the continuum, see \cite{Schroer:1963}.

\begin{center}
\textbf{Acknowledgment}
\end{center}

The author thanks K. Fredenhagen and O. G. Smolyanov for helpful discussions.

\appendix

\section{Norm estimates of observables\label{estimates}}

In the following $P_{S}\left( \Delta \right) $ with intervals $\Delta
\subset \mathbb{R}$ denotes the spectral family of the potential $F$. Let $%
\Delta _{1}$ and $\Delta _{2}$ be closed intervals of the real axes, and let 
$(\alpha ,\beta )\in \Delta _{1}\times \Delta _{2}\subset \mathbb{R}%
^{2}\rightarrow \chi (\alpha ,\beta )\in \mathbb{C}$ be a differentiable
function with the uniform bounds $\left| \chi (\alpha ,\beta )\right| \leq
c_{1}$ and $\left| \frac{\partial }{\partial \beta }\chi (\alpha ,\beta
)\right| \leq c_{2}$. Then $\beta \in \Delta _{2}\rightarrow T_{2}(\beta
)=\int_{\Delta _{1}}\chi (\alpha ,\beta )P_{S}(d\alpha )\in \mathcal{B}(%
\mathcal{H}_{S})$ is a differentiable family of operators with the norm
estimates $\left\| T_{2}(\beta )\right\| \leq c_{1}$ and $\left\|
T_{2}^{\prime }(\beta )\right\| \leq c_{2}$. If $A\in \mathcal{B}(\mathcal{H}%
_{S})$ is a bounded operator, the function $\beta \in \Delta _{2}\rightarrow
T(\beta )=T_{2}(\beta )A\in \mathcal{B}(\mathcal{H}_{S})$ is again
differentiable with the uniform estimates 
\begin{equation}
\left\| T(\beta )\right\| \leq c_{1}\left\| A\right\| \;\mathrm{and}%
\;\left\| T^{\prime }(\beta )\right\| \leq c_{2}\left\| A\right\|
\label{e.1}
\end{equation}
For all intervals $\Delta _{2}$ the Stieltjes integrals $\int_{\Delta
_{2}}T(\beta )P_{S}(d\beta )$ are well defined. Let $\Delta _{2}=\left[ a,b%
\right] $ be an interval of finite length. Then partial integration yields
the operator identity $\int_{\Delta _{2}}T(\beta )P_{S}(d\beta
)=T(b)E(b)-T(a)E(a)-\int_{\Delta _{2}}T^{\prime }(\beta )E(\beta )d\beta $
with the projection operators $E(\beta ):=P_{S}\left( (-\infty ,\beta
])\right) $, and the inequalities (\ref{e.1}) imply the bound 
\begin{equation}
\left\| \int_{\Delta _{2}}T(\beta )P_{S}(d\beta )\right\| \leq \left(
2c_{1}+\left| \Delta _{2}\right| c_{2}\right) \left\| A\right\| .
\label{e.3}
\end{equation}

The norm of $P_{S}(\Delta _{1})A(t)P_{S}(\Delta _{2})$, where $A(t)$ is the
Heisenberg operator (\ref{mod.10}), can now be estimated using (\ref{e.3}).
If $\Delta _{1}$ and $\Delta _{2}$ are disjoint intervals with a distance $%
\delta $, the constants $c_{1}$ and $c_{2}$ have to be substituted by the
upper bounds in (\ref{mod.12}), i.e. $c_{1}=c_{2}=c\left( 1+\delta ^{2}\psi
(t)\right) ^{-\gamma }$.

\section{The van Hove model\label{vanHove}}

Let $F\circ G$ denote the symmetric tensor product of the Fock space $%
\mathcal{F}(\mathcal{H}_{1})$ with vacuum $1_{vac}$. For all $f\in \mathcal{H%
}_{1}$ the exponential vectors $\exp f=1_{vac}+f+\frac{1}{2}f\circ f+...$
converge within $\mathcal{F}(\mathcal{H}_{1})$, the inner product being $%
\left( \exp f\mid \exp g\right) =\exp \left( f\mid g\right) $. The linear
span of all exponential vectors $\left\{ \exp f\mid f\in \mathcal{H}%
_{1}\right\} $ is dense in $\mathcal{F}(\mathcal{H}_{1})$. The creation
operators $a^{+}(f)$ are uniquely determined by $a^{+}(f)\exp g=f\circ \exp
g=\frac{\partial }{\partial \lambda }\exp (f+\lambda g)\mid _{\lambda
=0},\;f,g\in \mathcal{H}_{1}$ and the annihilation operators are given by $%
a(g)\exp f=\left( g\mid f\right) \exp f$. These operators satisfy the
standard commutation relations $\left[ a(f),a^{+}(g)\right] =\left( f\mid
g\right) $. If $M$ is a operator on $\mathcal{H}_{1}$ then $\Gamma (M)$ is
uniquely defined as operator on $\mathcal{F}(\mathcal{H}_{1})$ by $\Gamma
(M)\exp f:=\exp (Mf)$, and the derivation $d\Gamma (M)$ is defined by $%
d\Gamma (M)\exp f:=(Mf)\circ \exp f$.

As explicit example we may take $\mathcal{H}_{1}=\mathcal{L}^{2}(\mathbb{R}%
^{n}) $ with inner product \newline
$\left( f\mid g\right) =\int_{\mathbb{R}^{n}}\overline{f(k)}g(k)d^{n}k$. The
one-particle Hamilton operator can be chosen as \newline
$\left( Mf\right) (k):=\varepsilon (k)f(k)$ with the positive energy
function $\varepsilon (k)=c\left| k\right| ,\,c>0,\,k\in \mathbb{R}^{n}.$ Let $%
a_{k}^{\#},\,k\in \mathbb{R}^{n}$, denote the distributional
creation/annihilation operators, such that $a^{+}(f)=\int
a_{k}^{+}\,f(k)d^{n}k$ and $a(f)=\int a_{k}\,\overline{f(k)}d^{n}k$, then
the Hamiltonian $H_{E}=d\Gamma (M)$ coincides with $H_{E}=\int \varepsilon
(k)a_{k}^{+}a_{k}d^{n}k$.

For arbitrary elements $g\in \mathcal{H}_{1}$ the unitary Weyl operators are
defined on the set of exponential vectors by $T(g)\exp f=\mathrm{e}^{-\left(
g\mid f\right) -\frac{1}{2}\left\| g\right\| ^{2}}\exp (f+g)$. This
definition is equivalent to $T(g)=\exp \left( a^{+}(g)-a(g)\right) $. The
Weyl operators are characterized by the properties 
\begin{equation}
\begin{array}{c}
T(g_{1})T(g_{2})=T(g_{1}+g_{2})\,\exp \left( -i\mathrm{Im}\left( g_{1}\mid
g_{2}\right) \right) \\ 
\left( 1_{vac}\mid T(g)1_{vac}\right) =\exp \left( -\frac{1}{2}\left\|
g\right\| ^{2}\right) .
\end{array}
\label{f.4}
\end{equation}
The time evolution on the Fock space is given by $U(t)=\exp
(-iH_{E}t)=\Gamma \left( V(t)\right) $ with $V(t):=\exp (-iMt)$. For
exponential vectors we obtain $U(t)\exp f=\exp \left( V(t)f\right) $. From
these equations the dynamics of the Weyl operators follows as 
\begin{equation}
U^{+}(t)T(g)U(t)=T\left( V^{+}(t)\,g\right) .  \label{f.6}
\end{equation}
For fixed $h\in \mathcal{H}_{1}$ the unitary operators $T^{+}(h)U(t)T(h),\,t%
\in \mathbb{R}$, form a one parameter group which acts on exponential vectors as%
\newline
$T^{+}(h)U(t)T(h)\exp f=\exp \left( \left( h\mid V(t)(f+h)-f\right) -\left\|
h\right\| ^{2}\right) \exp \left( V(t)(f+h)-h\right) $. For \newline
$h\in \mathcal{H}_{1}$ with $Mh\in \mathcal{H}_{1}$ the generator of this
group is easily identified with $T^{+}(h)H_{E}T(h)=\,H_{E}+\Phi (Mh)+\left(
h\mid Mh\right) $, where $\Phi (.)$ is the field operator. This identity was
first derived by Cook \cite{Cook:1961} by quite different methods. If $h$
satisfies $M^{-1}h\in \mathcal{H}_{1}$ we obtain 
\begin{equation}
T^{+}(M^{-1}h)H_{E}T(M^{-1}h)-\left\| M^{-\frac{1}{2}}h\right\|
^{2}=H_{E}+\Phi (h)  \label{f.9}
\end{equation}
which is the Hamiltonian of the van Hove model \cite{Hove:1952}, see also, 
\cite{Berezin:1966} p.166ff, and \cite{Emch:1972}.

For all $h\in \mathcal{H}_{E}$ with $M^{-\frac{1}{2}}h\in \mathcal{H}_{E}$
the field operator $\Phi (h)$ satisfies the estimate 
\begin{equation}
\left\| \Phi (h)\psi \right\| \leq 2\left\| M^{-\frac{1}{2}}h\right\|
\left\| \sqrt{H_{E}}\psi \right\| +\left\| h\right\| \left\| \psi \right\| ,
\label{f.10}
\end{equation}
where $\psi \in \mathcal{F}(\mathcal{H}_{1})$ is an arbitrary vector in the
domain of $H_{E}$, see e.g. eq. (2.3) of \cite{Arai/Hirokawa:1997}. As
consequences we obtain

\begin{lemma}
\label{Selfadjoint}The operators $H_{E}+\lambda \Phi (h),\,\lambda \in \mathbb{R%
}$, are self-adjoint on the domain of $H_{E}$ if $h\in \mathcal{H}_{1}$ and $%
M^{-\frac{1}{2}}h\in \mathcal{H}_{1}$. The operator $H_{E}-\frac{1}{2}\Phi
^{2}(h)$ has the lower bound $H_{E}-\frac{1}{2}\Phi ^{2}(h)\geq -\left\|
h\right\| ^{2}$, if $h\in \mathcal{H}_{1}$ and $\left\| M^{-\frac{1}{2}%
}h\right\| \leq 2^{-1}$.
\end{lemma}

\begin{proof}
From (\ref{f.10}) and the numerical inequality $\sqrt{x}\leq ax+(4a)^{-1}$,
valid for $x\geq 0$ and $a>0$, we obtain a bound $\left\| \Phi (h)\psi
\right\| \leq c_{1}\left\| H_{E}\psi \right\| +c_{2}\left\| \psi \right\| $
with positive numbers $c_{1},\,c_{2}>0$ where $c_{1}$ can be chosen
arbitrarily small. Then the Kato-Rellich Theorem yields the first statement.

\noindent From (\ref{f.10}) we obtain\newline
$\left\| \Phi (h)\psi \right\| ^{2}\leq 4\left\| M^{-\frac{1}{2}}h\right\|
^{2}\left( \psi \mid H_{E}\psi \right) +4\left\| M^{-\frac{1}{2}}h\right\|
\left\| h\right\| \left\| \sqrt{H_{E}}\psi \right\| \left\| \psi \right\|
+\left\| h\right\| ^{2}\left\| \psi \right\| ^{2}\newline
\leq 8\left\| M^{-\frac{1}{2}}h\right\| ^{2}\left( \psi \mid H_{E}\psi
\right) +2\left\| h\right\| ^{2}\left\| \psi \right\| ^{2}.$ Hence the
operator inequalities \newline
$0\leq \frac{1}{2}\Phi ^{2}(h)\leq 4\left\| M^{-\frac{1}{2}}h\right\|
^{2}H_{E}+\left\| h\right\| ^{2}I_{E}$ hold, and we have derived the second
statement.
\end{proof}

Therefore the total Hamiltonian (\ref{mod.3}) is semibounded, and the
unitary operators \newline
$U_{\lambda }(t)=\exp \left( -i(H_{E}+\lambda \Phi (h))t\right) $ are well
defined if (\ref{field.4}) is satisfied.

In a first step we evaluate the expectation value of (\ref{field.5}) $%
U_{\alpha \beta }(t)=U_{\alpha }(-t)U_{\beta }(t)$ for a coherent state (=
normalized exponential vector) $\exp \left( f-\frac{1}{2}\left\| f\right\|
^{2}\right) $ under the additional constraint $M^{-1}h\in \mathcal{H}_{1}$.
This assumption allows to use the identity (\ref{f.9}) which reduces all
calculations to the Weyl relations and the vacuum expectation (\ref{f.4}).
The extension to the general case, which violates $M^{-1}h\in \mathcal{H}%
_{1} $, can then be performed by a continuity argument.

If $M^{-1}h\in \mathcal{H}_{1}$ the identity (\ref{f.9}) implies\newline
$U_{\lambda }(t)=T(-\lambda M^{-1}h)U_{0}(t)T(\lambda M^{-1}h)\exp \left(
i\lambda ^{2}\left( h\mid M^{-1}h\right) t\right) $. Then $U_{\alpha \beta
}(t)=U_{\alpha }(-t)U_{\beta }(t)$ can be evaluated with the help of (\ref
{f.4}) and (\ref{f.6}) with the result 
\begin{equation}
\begin{array}{l}
U_{\alpha \beta }(t)=T\left( (\alpha -\beta )\left( V^{+}(t)-I\right)
M^{-1}h\right) \,\exp \left( -i\varphi _{1}(t)\right) , \\ 
\varphi _{1}(t)=(\alpha ^{2}-\beta ^{2})\left\{ \left( h\mid M^{-1}h\right)
t+\left( M^{-1}h\mid M^{-1}\sin (Mt)h\right) \right\} .
\end{array}
\label{w.2}
\end{equation}
Let $\omega (f)$ denote the projection operator onto the normalized coherent
state \newline
$\exp \left( f-\frac{1}{2}\left\| f\right\| ^{2}\right) ,f\in \mathcal{H}%
_{1} $, then $\mathrm{tr}_{E}U_{\alpha \beta }(t)\omega (f)$ is evaluated as 
\newline
$\left( 1_{vac}\mid T^{+}(f)U_{\alpha \beta }(t)T(f)1_{vac}\right) =\left(
1_{vac}\mid T\left( (\alpha -\beta )\left( V^{+}(t)-I\right) M^{-1}h\right)
1_{vac}\right) \exp \left( -i\varphi (t)\right) $ \newline
with the phase \newline
$\varphi (t)=2(\alpha -\beta )\,\mathrm{Im}\left( f\mid \left(
I-V^{+}(t)\right) M^{-1}h\right) +(\alpha ^{2}-\beta ^{2})\left( \left(
M^{-1}h\mid ht+M^{-1}\sin (Mt)h\right) \right) $. Using the second identity
of (\ref{f.4}) we finally obtain 
\begin{equation}
\mathrm{tr}_{E}U_{\alpha \beta }(t)\omega (f)=\exp \left( -\frac{\left(
\alpha -\beta \right) ^{2}}{2}\left\| \left( V^{+}(t)-I\right)
M^{-1}h\right\| ^{2}\right) \exp \left( -i\varphi \right) .  \label{w.5}
\end{equation}
Under the assumption $M^{-1}h\in \mathcal{H}_{1}$ the norm $\left\| \left(
V^{+}(t)-I\right) M^{-1}h\right\| $ is uniformly bounded in $t$ and the
trace (\ref{w.5}) does not vanish for $t\rightarrow \infty $. But
nevertheless one can achieve a strong decrease which persists for some
finite time interval. This period can be chosen arbitrarily large if the low
energy contributions are strong; but inevitably, recurrences exist \cite
{Kupsch:2000}.

For vectors $h\in \mathcal{H}_{1}$ with $M^{-\frac{1}{2}}h\in \mathcal{H}_{1}
$ but $M^{-1}h\in \mathcal{H}_{1}$ we first prove that $\mathrm{tr}%
_{E}U_{\alpha \beta }(t)\omega (f)$ is again given by the identity (\ref{w.5}%
). Then we derive the essential statement that the norm $\left\| \left(
V^{+}(t)-I\right) M^{-1}h\right\| $ diverges for $t\rightarrow \infty $, and
consequently superselection sectors are induced for all $\alpha \neq \beta $.

The operators $H_{E}+\lambda \Phi (h)$ are self-adjoint on the domain of $%
H_{E}$ if $h\in \mathcal{H}_{1}$ and $M^{-\frac{1}{2}}h\in \mathcal{H}_{1}$.
Therefore it is possible to extend the result (\ref{w.5}) to Hamilton
operators which satisfy these constraints but violate $M^{-1}h\in \mathcal{H}%
_{1}$. To make this statement more explicit we introduce the norm 
\begin{equation}
\left| \left\| h\right\| \right| :=\left\| h\right\| +\left\| M^{-\frac{1}{2}%
}h\right\| .  \label{w.6}
\end{equation}
Let $h_{n}\in \mathcal{H}_{1},\,n=1,2,...,$ be a sequence of real vectors
which converges in this topology to a vector $h$, then we know from (\ref
{f.10}) and the proof of Lemma \ref{Selfadjoint} that there exist two null
sequences of positive numbers $c_{1n}$ and $c_{2n}$ such that 
\[
\left\| \left( \Phi (h_{n})-\Phi (h)\right) \psi \right\| \leq c_{1n}\left\|
\left( H_{E}+\Phi (h)\right) \psi \right\| +c_{2n}\left\| \psi \right\| .
\]
Hence the operators $H_{E}+\Phi (h_{n})$ converge strongly to $H_{E}+\Phi (h)
$. Then Theorem 4.4 of \cite{Maslov:1972} or Theorem 3.17 of \cite
{Davies:1980} imply the strong convergence of $U(h_{n};t)=\exp \left(
-i\left( H_{E}+\Phi (h_{n})\right) t\right) $ to $U(h;t)=\exp \left(
-i\left( H_{E}+\Phi (h)\right) t\right) $, uniformly in all intervals $0\leq
t\leq s<\infty $. The operators $U_{\alpha \beta ,n}(t):=\exp \left( i\left(
H_{E}+\alpha \Phi (h_{n})\right) t\right) \exp \left( -i\left( H_{E}+\beta
\Phi (h_{n})\right) t\right) $ converge therefore in the weak operator
topology to $U_{\alpha \beta }(t)$. For $n=1,2,..$ we can calculate the
corresponding traces $\mathrm{tr}_{E}U_{\alpha \beta ,n}(t)\omega (f)$ with
the result (\ref{w.5}) where $h$ has to be substituted by $h_{n}$. Since (%
\ref{w.5}) is continuous in the variable $h$ in the topology (\ref{w.6}) the
limit for $n\rightarrow \infty $ is again given by (\ref{w.5}).

To derive the divergence of $\left\| \left( V^{+}(t)-I\right)
M^{-1}h\right\| $ for $t\rightarrow \infty $ we introduce the spectral
resolution $P_{M}(d\lambda )$ of the one-particle Hamilton operator $M$. The
energy distribution of the vector $h\in \mathcal{H}_{1}$ is given by the
measure $d\sigma _{h}(\lambda )=\left( h\mid P_{M}(d\lambda )h\right) $. The
norm of $\left( V^{+}(t)-I\right) M^{-1}h$ is the square root of 
\begin{equation}
\psi (t):=\left\| \left( V^{+}(t)-I\right) M^{-1}h\right\| ^{2}=4\int_{\mathbb{R%
}_{+}}\lambda ^{-2}\sin ^{2}\frac{\lambda t}{2}\,d\sigma _{h}(\lambda ).
\label{d.2}
\end{equation}
This integral is well defined for all $h\in \mathcal{H}_{1}$, and $\psi (t)$
is differentiable for $t\in \mathbb{R}$.

\begin{lemma}
If $M^{-1}h\notin \mathcal{H}_{1}$, i.e. 
\begin{equation}
\int_{\varepsilon }^{\infty }\lambda ^{-2}\,d\sigma _{h}(\lambda )\nearrow
\infty \quad \mathrm{if}\;\varepsilon \rightarrow +0,  \label{d.3}
\end{equation}
then the integral (\ref{d.2}) diverges for $t\rightarrow \infty $.
\end{lemma}

\begin{proof}
Since the operator $M$ has an absolutely continuous spectrum, the measure $%
d\sigma _{h}(\lambda )$ is absolutely continuous with respect to the
Lebesgue measure $d\lambda $ on $\mathbb{R}_{+}$. Consequently, the measure $%
\lambda ^{-2}\,d\sigma _{h}(\lambda )$ is absolutely continuous with respect
to the Lebesgue measure on any interval $\left( \varepsilon ,\infty \right) $
with $\varepsilon >0$. The identity $\sin ^{2}\frac{\lambda t}{2}=\frac{1}{2}%
\left( 1-\cos \lambda t\right) $ and the Lebesgue Lemma therefore imply $%
\lim_{t\rightarrow \infty }\int_{\varepsilon }^{\infty }\lambda ^{-2}\sin
^{2}\frac{\lambda t}{2}\,d\sigma _{h}(\lambda )=\frac{1}{2}\int_{\varepsilon
}^{\infty }\lambda ^{-2}\,d\sigma _{h}(\lambda )$. Given a number $N>0$ the
assumption (\ref{d.3}) yields the existence of an $\varepsilon >0$ such that 
\begin{equation}
\lim_{t\rightarrow \infty }\int_{\varepsilon }^{\infty }\lambda ^{-2}\sin
^{2}\frac{\lambda t}{2}\,d\sigma _{h}(\lambda )=\frac{1}{2}\int_{\varepsilon
}^{\infty }\lambda ^{-2}\,d\sigma _{h}(\lambda )>N.  \label{d.4}
\end{equation}
From the inequality $\int_{\mathbb{R}_{+}}\lambda ^{-2}\sin ^{2}\frac{\lambda t%
}{2}\,d\sigma _{h}(\lambda )\geq \int_{\varepsilon }^{\infty }\lambda
^{-2}\sin ^{2}\frac{\lambda t}{2}\,d\sigma _{h}(\lambda )$ we then obtain%
\newline
$\int_{0}^{\infty }\lambda ^{-2}\sin ^{2}\frac{\lambda t}{2}\,d\sigma
_{h}(\lambda )>N$ for sufficiently large $t$. Since the number $N$ can be
arbitrarily large the integral (\ref{d.2}) diverges for $t\rightarrow \infty 
$.
\end{proof}

If $d\sigma _{h}(\lambda )$ satisfies additional regularity conditions, we
can obtain uniform estimates of the divergence. E. g. $d\sigma _{h}(\lambda
)\cong c\cdot \lambda ^{2\mu }d\lambda $ with $0<\mu <\frac{1}{2}$ and $c>0$
in a neighbourhood of $\lambda =+0$ implies a powerlike divergence $\psi
(t)\sim t^{1-2\mu }$.

So far the reference state $\omega $ has been a coherent state. But the
results remain obviously true if the reference state is a finite linear
combination of coherent states or a finite mixture of coherent states.

As a final remark we indicate a modification of the model, which does not
use the absolute continuity of the spectrum of $M$. But we still need a
dominating low energy contribution in the interaction. More precisely, we
assume that $\sigma _{h}(\lambda )\equiv \int_{0}^{\lambda }d\sigma
_{h}(\alpha )$ behaves at low energies like 
\begin{equation}
\lambda ^{-2}\sigma _{h}(\lambda )\nearrow \infty \;\mathrm{if}\;\lambda
\rightarrow +0.  \label{d.5}
\end{equation}
Then we can derive the divergence of (\ref{d.2}) by the inequalities\newline
$\psi (t)\geq 4\int_{\mathbf{0}}^{\frac{\pi }{t}}\lambda ^{-2}\sin ^{2}\frac{%
\lambda t}{2}\,d\sigma _{h}(\lambda )\geq \frac{4}{\pi ^{2}}t^{2}\int_{%
\mathbf{0}}^{\frac{\pi }{t}}\,d\sigma _{h}(\lambda )=\frac{4}{\pi ^{2}}%
t^{2}\,\sigma _{h}(\frac{\pi }{t})$ using $\sin x\geq \frac{2}{\pi }x$ if $%
0\leq x\leq \frac{\pi }{2}$. For measures $d\sigma _{h}(\lambda )\sim
\lambda ^{2\mu }d\lambda $ the assumption (\ref{d.5}) is more restrictive
than (\ref{d.3}) -- it excludes $d\sigma _{h}(\lambda )\sim \lambda d\lambda 
$ which satisfies the conditions of Lemma 2. But (\ref{d.5}) is also
meaningful for point measures $d\sigma _{h}(\lambda )$, and $M$ may be an
operator with a pure point spectrum. The Boson field can therefore be
substituted by an infinite family of harmonic oscillators, which have zero
as accumulation point of their frequencies. Such an example has been
discussed -- also for KMS states -- by Primas \cite{Primas:2000}.


\begin{thebibliography}{10}

\bibitem{Arai/Hirokawa:1997}
A.~Arai and M.~Hirokawa.
\newblock On the existence and uniqueness of ground states of a generalized
  spin-boson model.
\newblock {\em J. Funct. Anal.}, 151:455--503, 1997.

\bibitem{Araki:1980}
H.~Araki.
\newblock {A remark on Machida-Namiki theory of measurement}.
\newblock {\em Prog. Theor. Phys.}, 64:719--730, 1980.

\bibitem{Bell:1975}
J.~S. Bell.
\newblock {On wave packet reduction in the Coleman-Hepp model}.
\newblock {\em Helv. Phys. Acta}, 48:93--98, 1975.
\newblock {Reprinted in J. S. Bell: Speakable and Unspeakable in Quantum
  Mechanics, CUP 1987}.

\bibitem{Berezin:1966}
F.~A. Berezin.
\newblock {\em The Method of Second Quantization}.
\newblock Academic Press, New York, 1966.

\bibitem{Cook:1961}
J.~M. Cook.
\newblock {Asymptotic properties of a Boson field with given source}.
\newblock {\em J. Math. Phys.}, 2:33--45, 1961.

\bibitem{Davies:1980}
E.~B. Davies.
\newblock {\em One-Parameter Semigroups}.
\newblock Academic Press, London, 1980.

\bibitem{Emch:1972}
G.~G. Emch.
\newblock {\em {Algebraic Methods in Statistical Mechanics and Quantum Field
  Theory}}.
\newblock Wiley-Interscience, New York, 1972.

\bibitem{Emch:1972a}
G.~G. Emch.
\newblock On quantum measurement processes.
\newblock {\em Helv. Phys. Acta}, 45:1049--1056, 1972.

\bibitem{GJKKSZ:1996}
D.~Giulini, E.~Joos, C.~Kiefer, J.~Kupsch, I.~O. Stamatescu, and H.~D. Zeh.
\newblock {\em {Decoherence and the Appearance of a Classical World in Quantum
  Theory}}.
\newblock Springer, Berlin, 1996.

\bibitem{Hepp:1972}
K.~Hepp.
\newblock Quantum theory of measurement and macroscopic observables.
\newblock {\em Helv. Phys. Acta}, 45:236--248, 1972.

\bibitem{Hove:1952}
L.~{van} Hove.
\newblock Les difficult{\'e}s de divergences pour un mod{\`e}le particulier de
  champ quantifi{\'e}.
\newblock {\em Physica}, 18:145--159, 1952.

\bibitem{Joos/Zeh:1985}
E.~Joos and H.~D. Zeh.
\newblock The emergence of classical properties through interaction with the
  environment.
\newblock {\em Z. Phys.}, B59:223--243, 1985.

\bibitem{Kupsch:1998}
J.~Kupsch.
\newblock The structure of the quantum mechanical state space and induced
  superselection rules.
\newblock {\em Pramana - J. Phys.}, 51(5):615--624, 1998.
\newblock {quant-ph/9612033}.

\bibitem{Kupsch:2000}
J.~Kupsch.
\newblock Mathematical aspects of decoherence.
\newblock In Ph. Blanchard, D.~Giulini, E.~Joos, C.~Kiefer, and I.-O.
  Stamatescu, editors, {\em Decoherence: Theoretical,
  Experimental, and Conceptual Problems}, Lecture Notes in Physics 538, pages
  125--136, Springer, Berlin, 2000.

\bibitem{Maslov:1972}
V.~P. Maslov.
\newblock {\em {Th{\'e}orie des Perturbations et M{\'e}thodes Asymptotiques}}.
\newblock {{\'E}tudes Mathematiques}. Dunod, Paris, 1972.

\bibitem{Piron:1969}
C.~Piron.
\newblock Les r{\'e}gles de supers{\'e}lection continues.
\newblock {\em Helv. Phys. Acta}, 42:330--338, 1969.

\bibitem{Primas:2000}
H.~Primas.
\newblock Asymptotically disjoint quantum states.
\newblock In Ph. Blanchard, D.~Giulini, E.~Joos, C.~Kiefer, and I.-O.
  Stamatescu, editors, {\em Decoherence: Theoretical, Experimental, and
  Conceptual Problems}, Lecture Notes in Physics 538, pages 161--178, Springer,
  Berlin, 2000.

\bibitem{Reed/Simon:1975}
M.~Reed and B.~Simon.
\newblock {\em {Methods of Modern Mathematical Physics II: Fourier Analysis,
  Self-Adjointness}}.
\newblock Academic Press, New York, 1975.

\bibitem{Schroer:1963}
B~Schroer.
\newblock {Infrateilchen in der Quantenfeldtheorie}.
\newblock {\em Fortschr. Physik}, 11:1--32, 1963.

\bibitem{Wightman:1995}
A.~S. Wightman.
\newblock Superselection rules; old and new.
\newblock {\em Nuovo Cimento}, 110B:751--769, 1995.

\bibitem{Zurek:1982}
W.~H. Zurek.
\newblock Environment induced superselection rules.
\newblock {\em Phys. Rev.}, D26:1862--1880, 1982.

\end{thebibliography}
\end{document}